\documentclass[amsmath,amssymb,twocolumn]{aastex63}

\usepackage{textcomp}
\usepackage{gensymb}
\usepackage[hang,flushmargin,stable]{footmisc} 
\usepackage{graphicx}
\usepackage{txfonts}
\usepackage{xcolor}
\usepackage{float}
\usepackage{hyperref}
\usepackage{lineno}
\newcommand{\orcid}[1]{\href{https://orcid.org/#1}{\includegraphics[width=8pt]{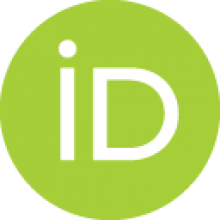}}}

\shorttitle{The orbital period changes for novae}
\shortauthors{Tang, Li \& Wang}
\begin{document}

\title{The orbital period changes for novae}

\author{Wenshi Tang\orcid{0000-0002-6588-9264}}
\affiliation{Department of Astronomy, Xiamen University, Xiamen 361005, China; \href{mailto:tangwenshi20@163.com}{tangwenshi20@163.com}}

\author{Xiang-Dong Li\orcid{0000-0002-6588-9264}}
\affiliation{School of Astronomy and Space Science, Nanjing University, Nanjing 210046, China; \href{mailto:lixd@nju.edu.cn}{lixd@nju.edu.cn}}
\affiliation{Key Laboratory of Modern Astronomy and Astrophysics (Nanjing University), Ministry of Education, Nanjing 210023, China}
\author{Bo Wang\orcid{https://orcid.org/0000-0002-3231-1167}}
\affiliation{Yunnan Observatories, Chinese Academy of Sciences, Kunming 650216, China}
\affiliation{International Centre of Supernovae, Yunnan Key Laboratory, Kunming, 650216, China}

\begin{abstract}

Cataclysmic variable (CVs) are close interacting binaries in which a white dwarf accretes materials from a low mass main sequence companion. CVs can experience nova eruptions due to low mass transfer rates. In the standard theory of CV evolution, the ejected materials during nova eruptions are assumed to leave the system in the form of fast, isotropic, optically thick winds, which predicts that novae only result in positive variation (expansion) of orbital period (i.e. positive $\Delta P$). In addition, the angular momentum losses (magnetic braking and gravitational radiation) only predicts a steady long-term decay in the orbital period of CVs, i.e. $\dot P$ is negative. Interestingly, an observation lasting over 30 years reveals positive and negative values for both $\Delta P$ and $\dot P$ in CVs, strongly conflicting with the standard evolutionary patterns. However, it cannot be excluded that these observations originate from short-term phenomena caused by nova eruptions because of a short timescale of observations. In this paper, we model the effect of instantaneous nova eruptions on the evolution of CVs, considering three mechanisms associated with mass loss in nova eruptions, including fast wind, Frank jet and binary-driven mass loss. By assuming that the observed $\Delta P$ and $\dot P$ are dominated by short-term phenomena, our results show that the binary-driven mass loss can explain almost all of the observations of normal CVs. However, the Frank jet may be needed for some of long-period CVs with evolved companions.
  
\end{abstract}


\section{Introduction}\label{Sect_Intro}
Cataclysmic variables (CVs) are short-period semi-detached binaries consisting of a mass-accreting white dwarf (WD) and a low-mass main sequence (MS) companion star. Instead of burning stably, the accreted materials by the WD firstly accumulate on the surface of the WD because of a low mass accretion rate ($10^{-10}-10^{-8}\,M_{\rm \odot}\rm yr^{-1}$, \citealt{2013ApJ...777..136W,wang18,2022MNRAS.510.6110P}). After a critical mass is accreted, unstable (runaway) nuclear burning occurs on the surface of the WD, which increases the luminosity of the system, called nova \citep[see ][for a review]{2021ARA&A..59..391C}. CVs are important for our understanding of accretion physics and binary evolution \citep{1995cvs..book.....W}. CVs are aslo potential progenitor of type Ia supernovae and gravitational wave \citep{1985ApJ...291..136S,2012NewAR..56..122W,2023MNRAS.525L..50S}. The evolution of CVs is controlled by angular momentum losses that drive the evolution of orbital period, either slow and steady change ($\dot P$) in  quiescence or sudden variation across a nova eruption ($\Delta P$; \citealt{2023MNRAS.525..785S}). $\Delta P$ is caused by the mass ejection process during a nova, while $\dot P$ is dominated by angular momentum losses in quiescence that mainly include gravitational radiation (GR) and magnetic braking (MB). 

In the standard theory of CV evolution, the nova ejecta is expelled as fast, isotropic, optically thick winds \citep[FWs; ][]{1966MNRAS.132..317F,2001ApJ...563..958K}, in which $\Delta P$ is always positive due to  the mass loss from nova eruptions \citep{1986ApJ...311..163S}. The long-term evolution of CVs is governed by MB at $P\gtrsim 3\,\rm hr$ \citep{1972ApJ...171..565S, 1983ApJ...275..713R} and by GR at $P\lesssim 2\,\rm hr$ \citep{1959flme.book.....L}. Both mechanisms result in the decay of $P$, i.e. negative $\dot P$. Because the widely adopted MB law in the standard theory only depends on the orbital period \citep{1972ApJ...171..565S, 1983ApJ...275..713R} so that all CVs will quickly join onto a unique evolution track \citep{2011ApJS..194...28K}. To test the proposed model, \cite{2023MNRAS.525..785S, 2024ApJ...966..155S} measured $\Delta P$ for 14 novae and collected $\dot P$ for 52 CVs, revealing that both the measured $\Delta P$ and $\dot P$ have positive and negative values. This is in contradiction with the standard theory that only produces positive $\Delta P$ and negative $\dot P$. Even for negative $\dot P$, the standard MB obtains $\dot P$ smaller than observations by several orders. 
In addition, some recurrent novae, such as T CrB and U Sco, underwent highly significant changes of $\Delta P$ and $\dot P$ from eruption to eruption, which is impossible in the standard theory. Therefore, \cite{2024ApJ...966..155S} argued that the standard MB theory may be wrong by orders of magnitude. For this reason, it is of great importance to provide an explanation for observed $\Delta P$ and $\dot P$.

However, as discussed in \cite{2024arXiv240603948K}, the predicted $\dot P$ by MB is a mean effect over a timescale ($>10^{5}$\,yr) of which angular momentum loss moves the Roche lobe of the donor one density scaleheight. Therefore, it is likely to lead to misleading conclusions by using the observed $\dot P$ to constrain the MB theory, as the observations by Schaefer only span $\sim 30$\,yr.  It has been proposed that the reported period changes should originate from short–term phenomena \citep{2024arXiv240603948K}. The purpose of this paper is to investigate whether the observed $\Delta P$ and $\dot P$ for novae can be explained by short-term variations caused by nova eruptions from a view of self-consistent binary evolution. 

 The process of mass ejection during nova eruptions may be more complicated than the standard model. Some observed phenomena for classical novae deviate from the simple fast wind assumption, such as aspherical ejecta, multiple modes of mass ejection \citep{2014Natur.514..339C, 2020ApJ...905...62A}. \cite{2022ApJ...938...31S} carried out a hydrodynamical simulation of classical nova outflow. They found that most of ejecta during a nova eruption initially have a low velocity and are accelerated at radii beyond the WD’s Roche-lobe radius. This implies that the companion may play a critical role in driving mass ejection during nova eruptions, called the binary-driven mass loss (BDML). \cite{2024ApJ...977...34T} recently suggested that BDML can produce broad mass transfer rate and companion radius distributions to match better CV's properties. Meanwhile, \cite{2024ApJ...977...34T} found that BDML can explain some observed $\Delta P$ in novae (see Figure 7 in their paper). However, the theoretical $\Delta P$ in Figure 7 of that paper is a semi-analytical estimation that does not rely on detailed binary simulations. \footnote{In Figure 7 of \cite{2024ApJ...977...34T},  the theoretical $\Delta P$ is estimated by adopting white dwarf masses and nova ejecta masses from Table 10 of \cite{2023MNRAS.525..785S} and the inferred $f_{\rm ML,L2}$ from Figure 8 of \cite{2022ApJ...938...31S}. Then $\Delta P$ was calculated using Equation (17) of \cite{2024ApJ...977...34T}). This does not involve detailed binary simulations.} Moreover, \cite{2024ApJ...977...34T} did not focus on explaining the observed $\dot P$. Besides, to account for observed $\Delta P$, the jet mechanism is proposed by J. Frank in \cite{2020MNRAS.492.3343S}, which arises from asymmetric mass ejection in novae, named as the Frank jet. However, the Frank jet has not been tested through self-consistent binary evolution. In this paper, we will compare the effectiveness of fast wind, BDML and Frank jet in explaining the observed $\Delta P$ and $\dot P$ by incorporating them into detailed binary evolution code. 

This paper is organized as follow. In Section\,\ref{Sect_method}, we describe our methods. In Section\,\ref{Sect-results}, we present our results and compare with observations. Lastly, summary and discussion are given in Section\,\ref{Sect-Con-Diss}.

\section{Assumptions and calculations}\label{Sect_method}

Similar to \cite{2024ApJ...977...34T}, we use the stellar evolution code {\tt MESA} \citep[version 11701; ][]{2011ApJS..192....3P, 2013ApJS..208....4P,2015ApJS..220...15P,2019ApJS..243...10P} to model the detailed evolution of WD+MS binaries considering the influence of nova eruptions. We refer the reader to \cite{2024ApJ...977...34T} for detailed treatments for modeling the influence of nova eruptions on CV evolution. Here, we provide a brief description. 

We simulate the evolution of a large number of WD+MS binaries to get the parameter space for the formation of CVs, in which the initial WD masses ($M_{\rm WD,i}$) are set to $0.6-1.2\,M_{\rm \odot}$ in steps of 0.2\,$M_{\rm \odot}$, initial donor masses ($M_{\rm 2,i}$) are $0.4-2.4\,M_{\rm \odot}$ in steps of 0.2\,$M_{\rm \odot}$ and initial orbital period ($P_{\rm i}$) are $0.2-3.0$\,days in steps of 0.1\,days. The companions have solar metallicity and the angular momentum losses include gravitational radiation\citep{1959flme.book.....L}, magnetic braking \citep[with $\gamma =3$; ][]{1972ApJ...171..565S, 1983ApJ...275..713R} and mass loss which will be described below.
 
 
The evolutionary outcome of a WD+MS system depends on WD mass and mass transfer rate $\dot M_{2}$. If the mass transfer rate exceeds the critical rate for stable hydrogen burning ($\dot M_{\rm stable}$), the transferred material may be retained on the surface of the WD. The mass growth rate of the WD is expressed as
\begin{equation}
\dot M_{\rm WD} =  \eta_{\rm H}\eta_{\rm He} |\dot M_{\rm 2}|
\end{equation}
where $\eta_{\rm H}$ and $\eta_{\rm He}$ are the mass accumulation efficiencies for hydrogen and helium burning, respectively. $\eta_{\rm He}$ is adopted from the result of \cite{2017A&A...604A..31W}. For $\eta_{\rm H}$, we take the same form  in  \cite{2010MNRAS.401.2729W},
 \begin{equation}
\eta _{\rm H}=\left\{
 \begin{array}{ll}
 \dot{M}_{\rm crit}/|\dot{M}_{\rm 2}|, & {\rm if}\ |\dot{M}_{\rm 2}|> \dot{M}_{\rm crit},\\
 1, & {\rm if}\ \dot{M}_{\rm crit}\geq |\dot{M}_{\rm 2}|\geq \dot{M}_{\rm stable},\\
\end{array}\right.
\end{equation}
$\dot M_{\rm stable}$ and $\dot M_{\rm crit}$ are taken from \cite{2013ApJ...778L..32M}. The material that cannot be retained by the WD leaves the binary in the form of continuous fast winds.

 On the contrary, a system can behaves as a CV and experience nova eruptions if $|\dot M_{2}|<\dot M_{\rm stable}$. For this region, difference with the traditional continuous wind picture, we deal with the mass ejection process in novae as an instantaneous process. Therefore, the time step should be set to resolve the nova as much as possible. In {\tt MESA}, the timestep is limited to $max\_timestep$. In our calculation, we set $max\_timestep = f_{\rm dt}\times \tau_{\rm rec}$ with $f_{\rm dt}$ being an adjustable coefficient between [0,1] and $\tau_{\rm rec}$ being recurrence time of novae. In fact, this setting uses the recurrence timescale of the nova in the current step to control the evolution of the next step. Therefore, it does not guarantee that all novae will be resolved, since the recurrence timescale of the next nova is not known in advance. Resolving all novae is indeed very challenging, as some have very short recurrence timescales, which can cause MESA to fail due to excessively small time steps. Nevertheless, our goal is to control the computational time step and resolve novae as much as possible, thus we simply adopt this setting.
 
 As discussed in the Appendix B of \cite{2024ApJ...977...34T}, in principle, a smaller $f_{\rm dt}$ yields more precise results. However, an excessively small $f_{\rm dt}$ would be extremely time-consuming and resource-intensive. Moreover, too small an $f_{\rm dt}$ can cause many calculations to fail to proceed normally, especially for systems with massive white dwarfs. Therefore, for systems with $M_{\rm WD,i}\ge 1.0\,M_{\odot}$, we adopt $f_{\rm dt}$ = 0.5, whereas for $M_{\rm WD,i} < 1.0\,M_{\odot}$, we carry out calculations with $f_{\rm dt}$ = 0.03, 0.1, and 0.5\footnote{This results in the number ($N$) of steps between two successive nova events ranging from 0 to several tens. For $f_{\rm dt}$=0.5, approximately 30\%-40\% of novae have $N > 6$ and about a few to a dozen percent of novae have $N =0$ (indicating that the nova is not resolved). For $f_{\rm dt}$=0.03 and 0.1, more than 90\% and 70\%-80\% of novae, respectively, have $N > 6$ and less than a few percent  of the nova events have $N = 0$. Our overall conclusions are not very sensitive to the choice of $f_{\rm dt}$.}. Smaller $f_{\rm dt}$ like 0.01 would make the calculation be nonconvergent for almost all of the systems because of too small a time step. Then for a specific system, we select the model with the smallest $f_{\rm dt}$ for which the calculation successfully converges as our adopted model. $\tau_{\rm rec}$ as a function of $\dot M_{2}$ and $M_{\rm WD}$ is adopted from Figure\,2 of \cite{2021ARA&A..59..391C} (see \citealt{2024ApJ...977...34T} for details). Then the nova ejecta mass $M_{\rm ejecta}=|\dot M_{2}|\times \tau_{\rm rec}$. In the nova region, we assume that the transferred masses from the donor will be firstly accreted by the WD. Then, a nova eruption occurs when the accreted mass by the WD reaches $M_{\rm ejecta}$ and the materials with a mass of $M_{\rm ejecta}$ are instantaneously ejected, which means that the WD has no net mass growth in the nova region.  
 
 \begin{figure*}[ht]
\centering
\includegraphics[width=12cm,height=6cm]{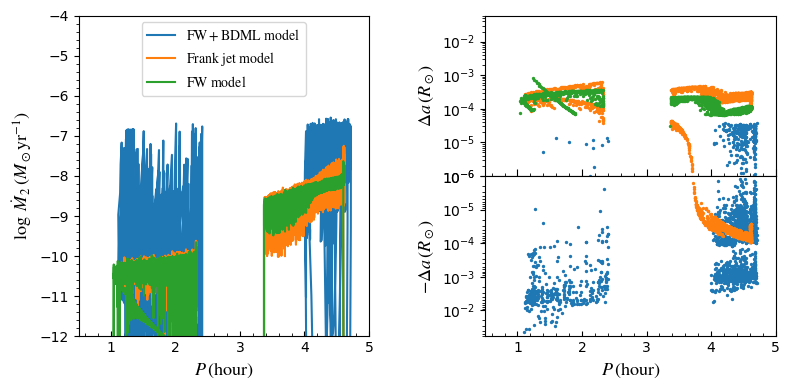}
\includegraphics[width=12cm,height=6cm]{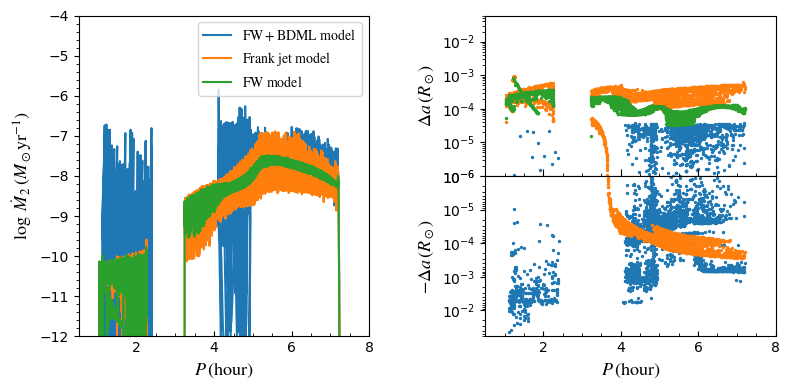}
 \caption{Examples of demonstrating the effects of BDML and jet on CV evolution, where the initial binary parameters are ($M_{\rm WD,i}, M_{\rm 2,i}, P_{\rm i}$)=($0.8\,M_{\rm \odot}, 0.6\,M_{\rm \odot}, 0.6\,\rm days$) and ($0.8\,M_{\rm \odot}, 1.0\,M_{\rm \odot}, 1.2\,\rm days$) for upper and lower panels, respectively. The left panels show the evolution of mass transfer rate as a function of orbital period, while the variation of binary separation across nova eruptions is shown in the right panels.  Note that the lowest mass transfer rate can be 0, while we only show a lower limit of log\,$\dot M_{2}=-12$ for clarity. The data colored by blue, orange and green represent the FW+BDML model, the Frank-jet model and the FW model, respectively. In the $\Delta a-P$ plane, the upper panel is positive $\Delta a$, while the lower panel is negative $\Delta a$. The green arc-like structure separated from the main sequence at $P<2\,\rm hr$ in $\Delta a-P$ plane corresponds to the track of part of period bouncer. For the system shown in upper panels, the FW+BDML model starts at a slightly longer period because the strong angular momentum loss due to BDML can cause expansion of the orbit to a value larger than that at the onset of mass transfer. Nevertheless, for the example shown in lower panels, the same start point of three models is attributed to a larger total mass, leading to a smaller impact from nova eruptions.}
\label{Fig-example}
\end{figure*}

Instantaneous mass ejection can cause a variation in binary separation $a$. We consider the mass ejection process associated nova eruptions as below. The orbital angular momentum of a binary is 
 \begin{equation}\label{Eq-Jorb}
 J_{\rm orb} = M_{\rm WD}M_{2}\sqrt{\frac{G a}{M}}.
 \end{equation}
 where $M=M_{\rm WD}+M_{2}$, and $G$ is the gravitational constant. We neglect any recapture of the nova ejecta by the companion star. Then the variation of binary separation induced by a nova is

\begin{equation}\label{Eq-delta-q}
 \frac{\Delta a}{a}=2\frac{\Delta J_{\rm orb}}{J_{\rm orb}}+\left(\frac{1+2q}{1+q}\right)\frac{M_{\rm ejecta}}{M_{\rm WD}}.
 \end{equation}
 where $q=M_{2}/M_{\rm WD}$. For mass loss, it may affect $\Delta a$ via three forms: fast wind, BDML and Frank jet. To  determine $\Delta a$, we construct three models: pure FW model, FW+BDML model and Frank-jet model.

 For the FW model, the angular momentum loss due to nova eruption is 
 \begin{equation}
    \Delta J_{\rm orb,FW}=-\left(\frac{M_{2}J_{\rm orb}}{MM_{\rm WD}}\right)M_{\rm ejecta}.
 \end{equation} 
 Then, 
\begin{equation}\label{Eq-Da-FW}
\frac{\Delta a_{\rm FW}}{a} = \left(\frac{M_{\rm ejecta}}{M_{\rm WD}}\right)\left(\frac{1}{1+q}\right).
\end{equation}
 \\
 
 For the FW+BDML model, we assume that a part of the lost materials during a nova eruption escapes from the system via fast winds (with a percentage of $f_{\rm ML, FW}$). The rest of materials with a fraction of $f_{\rm ML, L2}$ (=1-$f_{\rm ML, FW}$)  is lost driven by binary interaction and they finally leave the binary from the outer Lagrange point $L_{2}$ \citep{2022ApJ...938...31S}. The value of $f_{\rm ML, L2}$ as a function of $M_{\rm WD}$ and $M_{\rm ejecta}$ is obtained according to Fig.\,8 in \cite{2022ApJ...938...31S} (see \citealt{2024ApJ...977...34T} for details). The angular momentum loss associated mass loss from the $L_{\rm 2}$ point is $\Delta J_{{\rm orb},L_{2}}=-M_{\rm ejecta} a_{L_{2}}^{2}\omega$, where $\omega$ is the orbital angular velocity of the binary and $a_{L2}$ is the distance between the center of mass of the binary and the $L_{2}$ point (see Eqs.\,(14) and (15) in \citealt{2024ApJ...977...34T}). Then the change of the orbital angular momentum in a nova eruption is
 \begin{equation}\label{DJ-FW-L2}
\Delta J_{\rm orb} = f_{\rm ML, FW} \times \Delta J_{\rm orb, FW}+f_{\rm ML, L2}\times \Delta J_{\rm orb, L2}.
\end{equation}
 Combing Eqs.\,(\ref{Eq-delta-q}) and (\ref{DJ-FW-L2}), we can obtain $\Delta a$ for the FW+BDML model.

 For the Frank-jet model, all ejected masses flow away from the system via fast winds, meanwhile the lost masses impart a kick onto the WD because of asymmetric mass ejection \citep[i.e. Frank jet; ][]{2020MNRAS.492.3343S, 2023MNRAS.525..785S}. The variation of the orbital period caused by the jet is given by \cite{2023MNRAS.525..785S},
\begin{equation}\label{Dp-jet}
\frac{\Delta P_{\rm jet}}{P} = -1.5q\xi\left(\frac{M_{\rm ejecta}}{M_{\rm WD}+M_{2}}\right)\frac{V_{\rm ejecta}}{V_{\rm WD}}
\end{equation}
Here, $V_{\rm WD}$ is orbital velocity of the white dwarf. $V_{\rm ejecta}$ is velocity of nova ejecta, which is set to 1000\,$\rm km\,s^{-1}$ \citep{2023MNRAS.525..785S}. $\xi$ is a parameter describing the asymmetry of the ejecta, where $\xi$ = +1 represents the case where all the ejecta is in a hemisphere centred on the forward direction of the white dwarf’s orbital velocity, and -1 represents a backward facing hemisphere of ejecta.  To investigate the maximum effect of the jet, we let $\xi$ be 1 or -1 randomly in our calculation \footnote{ Although $|\xi|$ = 2 is theoretically possible which represents a narrow jet, we adopt $|\xi|$ = 1 as our fiducial limit as used in \cite{2023MNRAS.525..785S}. However, the results can be influenced by adopting $|\xi|$=2.}. The results with a continuous $\xi$ between 1 and -1 is located within the contour of $|\xi|=1$. Using Kepler's third law, Eq.\,(\ref{Dp-jet}) is transformed to $\Delta a_{\rm jet}$. Then the net variation of the separation is $\Delta a = \Delta a_{\rm FW}+\Delta a_{\rm jet}$.
 
\begin{figure*}[ht]
\centering
\includegraphics[width=7.5cm,height=7.5cm]{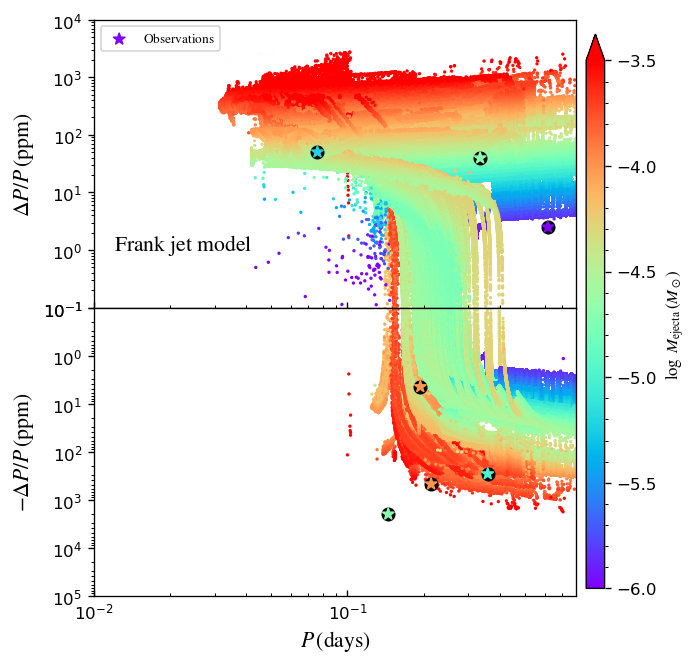}
\includegraphics[width=7.5cm,height=7.5cm]{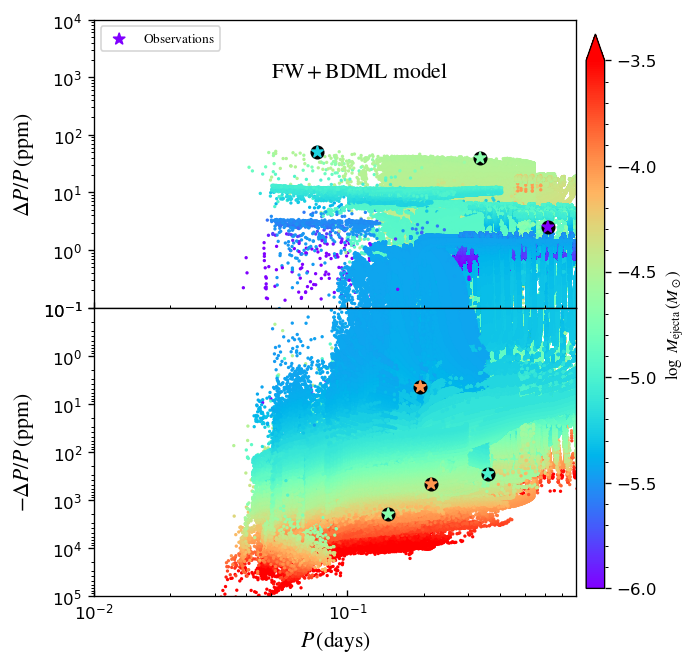}
 \caption{Comparison with observed $\Delta P/P$ in the unit of parts-per-million (ppm). The left and right panels are the Frank-jet model and the FW+BDML model, respectively. The colored stars are observed $\Delta P$ from \cite{2023MNRAS.525..785S}. The colored points are our evolutionary data which include all systems that can form CVs, where there are 433 and 389 individual tracks for the Frank-jet model and FW+BDML model, respectively. The upper panels show positive $\Delta P$, while the lower panels show negative $\Delta P$. The color bar represents the mass of nova ejecta, in which the nova ejecta masses for observed sources are taken from Table 10 of \cite{2023MNRAS.525..785S}.}
\label{Fig-Dp}
\end{figure*}

\begin{figure*}[ht]
\centering
\includegraphics[width=15cm,height=7.5cm]{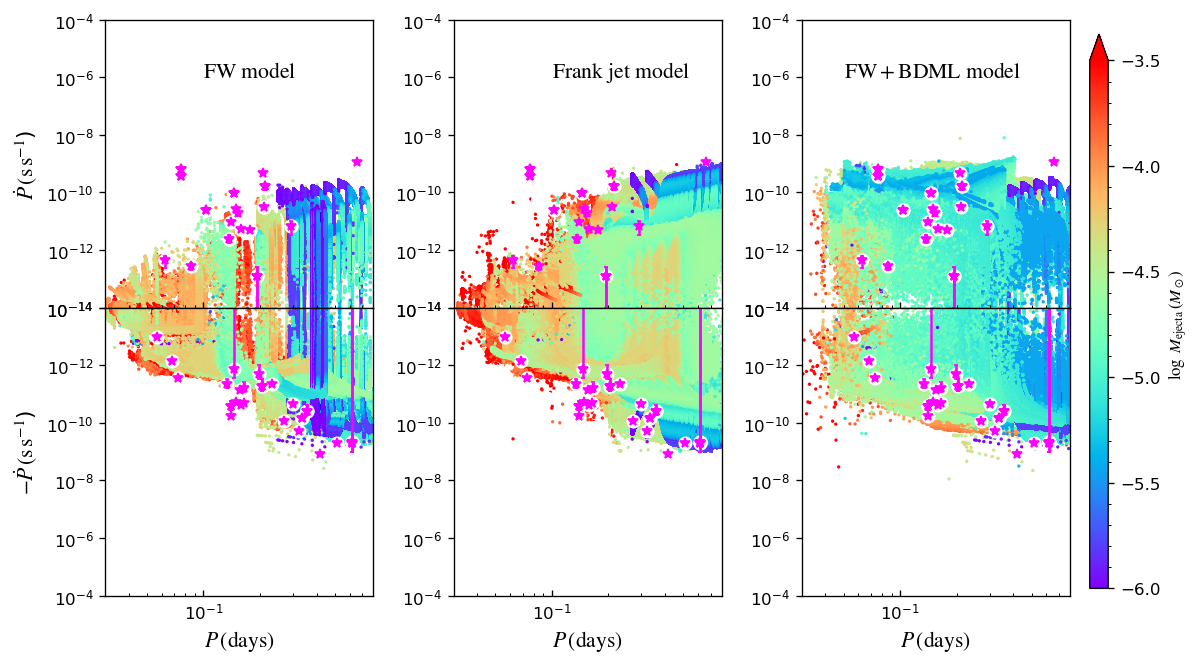}
 \caption{Comparison with observed $\dot P$ in the unit of s\,$\rm s^{-1}$. The left, middle and right panels are the FW model, Frank-jet model and the FW+BDML model, respectively. There are 384, 433 and 389 individual systems contributing to the FW model, Frank-jet model and FW+BDML model, respectively. The magenta stars are observed $\dot P$ from \cite{2024ApJ...966..155S}. The color bar represents the mass of nova ejecta. The upper panels show positive $\dot P$, while the lower panels show negative $\dot P$.}
\label{Fig-Pdot}
\end{figure*}

\section{Results}\label{Sect-results}
\subsection{Examples of the effects of BDML and jet on CV evolution}
Fig.\,\ref{Fig-example} shows two evolutionary examples demonstrating the effects of BDML and jet on CV evolution \footnote{Here we only present specific examples. The effect of BDML on the overall properties of CV population is given by \cite{2024ApJ...977...34T}, while the effect of jet on that will be presented by an upcoming paper.}. The upper panels and  lower panels depict the evolution of systems with initial parameters ($M_{\rm WD,i}, M_{\rm 2,i}, P_{\rm i}$) = ($0.8\,M_{\rm \odot}, 0.6\,M_{\rm \odot}, 0.6\,\rm days$) and ($0.8\,M_{\rm \odot}, 1.0\,M_{\rm \odot}, 1.2\,\rm days$), respectively. They show some interesting features.

The left panels show the evolution of mass transfer rate as a function of orbital period. We see that for the FW+BDML model, the mass transfer proceeds in a discontinuous manner. This is because that the strong angular momentum loss induced by BDML during nova eruptions can sustain a high mass transfer rate \footnote{In the FW+BDML model, $\lesssim$ 20–30\% of the mass is lost in the form of BDML, while the vast majority (over 70–80\%) is lost through fast winds. As a result, the system avoids undergoing runaway mass transfer.} and cause expansion of the orbit . This can be understood as follows. Large amounts of angular momentum loss in nova eruptions lead to an abrupt shrinkage of the orbit. Because the orbital variation is instantaneous, the companion cannot immediately adjust itself to the new orbit. The companion slightly overfills its Roche lobe and gives rise to an increase in the mass transfer rate. The rapid mass transfer has the effect of expanding the orbit since $q<1$ and the mass transfer timescale is shorter than the orbital decay time due to GR and MB (see also Section 3.1 in \citealt{2024ApJ...977...34T}). As the angular momentum-loss rate decreases after the nova eruptions, the companion can no longer maintain Roche-lobe overflow, leading to a detached phase (see also \citealt{2024ApJ...977...34T}). In contrast, for the FW model and Frank-jet model, the systems do not become completely detached, especially at long period, because the impact of nova eruptions on the orbit is less significant.

Notably, for the system with an initial more massive companion and longer orbital period (the lower panels), the FW+BDML model does not lead to detachment in the early stages of the evolution ($P\gtrsim 5$\,hr). This is attributed to a higher total mass of system at that stage, which results in a smaller impact from the novae. On the other hand, in the Frank-jet model, novae can lead to larger variation in mass transfer rate at longer period compared to shorter period above the period gap. This arises from the positive correlation between $\Delta a_{\rm jet}$ and orbital period (see Figure 9 in \citealt{2023MNRAS.525..785S}).

Compared with the FW model, the BDML has great impact on the evolution of the mass transfer rate across both long and short periods. The jet exerts a moderate impact at long periods, while it hardly affects the mass transfer rate at short periods. This stems from that $\Delta P_{\rm jet}$ decreases as decay of the orbital period. At $P\lesssim 2\,\rm hr$, $|\Delta a_{\rm jet}|$ becomes comparable to $\Delta a_{\rm FW}$ \citep{2023MNRAS.525..785S}.

In addition, the period gap is similar for the FW model and Frank-jet model, whereas the edges of the period gap are shifted to longer value  in the FW+BDML model. This is again a consequence of orbital expansion induced by BDML. However, in principle, the discrepancy in the period gap for the FW+BDML model may be alleviated by modifying the physics related to magnetic braking (see Appendix.\,\ref{Appenx1}), although this would require a more detailed investigation, which is beyond the scope of this paper. In any case, our purpose is to explore the possible mechanisms associated with mass loss process during nova eruptions in order to explain the observed $\dot P$ and $\Delta P$ of novae. This does not depend on the details of magnetic braking prescriptions and the exact location of the period gap.

The right panel shows the variation of binary separation across nova eruptions as a function of orbital period. The distribution of $\Delta a$ for the Frank-jet model splits up into two branches, which is the result of the choice of the value for $\xi$. The upper branch corresponds $\xi=+1$, while the lower branch is $\xi=-1$. 
The right panel shows that the FW model only produces positive $\Delta a$. Although the Frank-jet model and the FW+BDML model can produce both positive and negative $\Delta a$ at long periods, the Frank-jet model almost always results in positive values at short periods because the negative $\Delta a_{\rm jet}$ can be roughly offset by $\Delta a_{\rm FW}$ \citep{2023MNRAS.525..785S}.


\subsection{Comparison with observed $\Delta P$}
We compare our results with the observed $\Delta P$ in Fig.\,\ref{Fig-Dp}, where the colored points are our evolutionary data, including all systems that can form CVs. The figure displays 420 individual evolutionary tracks for the Frank-jet model and 379 tracks for the FW+BDML model. The FW model is not shown here because it only produces positive $\Delta P$ so that it is impossible to explain all observations. We see from Fig.\,\ref{Fig-Dp} that the Frank-jet model can cover most of observed sources. Note that although the prediction by the Frank-jet model is higher than the observation of the source with longest orbital period, a smaller $\xi$ can solve this discrepancy. However, the difficulty for the Frank-jet model is that it cannot explain the one with most negative $\Delta P$.

Instead, the FW+BDML model can account for all observations. For some sources, the nova ejecta masses predicted by our calculations are not fully consistent with the derived values by \cite{2023MNRAS.525..785S}, whereas $M_{\rm ejecta}$ provided by \cite{2023MNRAS.525..785S} is only approximate value with poor accuracy as noted by himself.

\subsection{Comparison with observed $\dot P$}
The steady orbital period changes in quiescence between novae can be written as \citep{2023MNRAS.525..785S}
\begin{equation}\label{Eq-Pdot1}
\dot P = \dot P_{\rm MT}+\dot P_{\rm GR}+\dot P_{\rm MB}+\dot P_{\rm \Delta P}
\end{equation}
where $\dot P_{\rm MT}$ is the contribution from the mass transfer.  $\dot P_{\rm GR}$ arises from gravitational radiation \citep{1959flme.book.....L} and $\dot P_{\rm MB}$ is from magnetic braking \citep{1972ApJ...171..565S,1983ApJ...275..713R}. These three terms are calculated by {\tt MESA} self-consistently \citep{2015ApJS..220...15P}.  The last term in above equation results from sudden mass loss process in novae and is expressed as $\dot P_{\rm \Delta P} = \Delta P/\tau_{\rm rec}$.  Although GR and MB always cause decay of orbit, the remaining two can lead to either orbital decay or expansion. As a result, the sign of  $\dot P$ is determined by their relative magnitudes.

Using our evolutionary data, the mean derivative of the orbital period is calculated as
\begin{equation}
\langle  \dot P \rangle= \frac{P_{\rm i}-P_{\rm i-1}}{\Delta t}
\end{equation}
where $P_{\rm i}$ and $P_{\rm i-1}$ denotes the orbital periods of $i$\,th and ($i-1$)\,th nova eruptions, respectively, and $\Delta t$ is their time interval. With such a formula, $\langle  \dot P \rangle$ includes all terms in Eq.\,(\ref{Eq-Pdot1}).

A comparison between our results and the observed $\dot P$ is shown in Fig.\,\ref{Fig-Pdot}. In this figure, there are 384, 433 and 389 individual evolutionary tracks contributing to the FW model, Frank-jet model and FW+BDML model, respectively\footnote{The difference in the numbers arises from the criterion used to determine the availability of a track. Ideally, we expect that every initial system can evolve up to the Hubble time. However, many simulations are prematurely terminated due to convergence issues in the {\tt MESA} code. We consider a track to be available if the system can evolve to a point where the companion star can lose at least 50\% of its initial mass. Although this criterion is somewhat arbitrary, it does not affect our conclusions.}. The period gap does not appear in these figures because some systems continue mass transfer while passing through it without developing any helium core. They are still included because they ultimately form CVs. Typically, these systems have relatively massive companions ($\gtrsim1.2\,M_{\rm \odot}$) and long initial orbital periods ($\gtrsim 1.8$\,days). However, they only contribute a small fraction of the overall CV population (see the population syntheses results of Figure 5 in \cite{2024ApJ...977...34T}. 

As expected, the FW model cannot explain all sources, consistent with the findings of \cite{2023MNRAS.525..785S}. Similarly, the Frank-jet model also fails to account for the complete set of observations, especially for short-period sources. In contrast, the FW+BDML model successfully explains nearly all observed data.

\section{Conclusion and Discussion}\label{Sect-Con-Diss}
In this paper, assuming that the observed $\Delta P$ and $\dot P$ in CVs arise from short-term phenomena, we explore the possible mechanisms associated with mass loss process of nova eruptions to account for the observation. The mass loss during nova eruptions may affect $\Delta P$ and $\dot P$ via three possible mechanisms: fast wind, Frank jet and binary-driven mass loss. By incorporating above three mechanisms into {\tt MESA} code, we carry out detailed binary evolution considering the influence of instantaneous nova eruptions . We find that both the FW model and the Frank-jet model can only explain part of the observations, while the FW+BDML model can cover nearly all observed $\Delta P$ and $\dot P$.

Some sources with extremely short ($P\lesssim 0.01$\,days) and long periods ($P\gtrsim 1$\,days) listed in \cite{2023MNRAS.525..785S,2024ApJ...966..155S} are not discussed in this paper. The companions of these sources are either WDs or sub-giants or red-giants, which are not included in our calculations. However, we argue that both the FW model and the FW+BDML model may be invalid for some of these sources. For example, the recurrent novae U Sco and T CrB have orbital periods of 1.23\,days and 227\,days, respectively. There are four and two period changes recorded by  \cite{2023MNRAS.525..785S,2024ApJ...966..155S}  for U Sco and T CrB, respectively.  Their $\Delta P$ and $\dot P$ vary drastically from eruption to eruption. In the FW model and the FW+BDML model, varying ejecta mass is the only way to obtain various $\Delta P$ and $\dot P$ for a specific system. However, $M_{\rm ejecta}$ for U Sco and T CrB seemingly have no significant variations from eruption to eruption \citep{2023MNRAS.525..785S}. For such systems, the jet mechanism may be required, which can obtain variable $\Delta P$ and $\dot P$ by changing $\xi$ (see Eq.\,(\ref{Dp-jet})). Furthermore, it is also possible that the three mechanisms work together, whereas it is difficult to determine their relative importance.

\nolinenumbers 
\section*{Acknowledgments}
\quad We are grateful to the referees for insightful comments that helped improve the manuscript. This work was supported by the Natural Science Foundation of China (Nos 12041301, 12121003 and 12225304), the National Key Research and Development Program of China (2021YFA0718500), the Western Light Project of CAS (No. XBZG-ZDSYS-202117), and the International Centre of Supernovae, Yunnan Key Laboratory (No. 202302AN360001).

\section*{Data Availability}
\quad All data underlying this article will be shared on reasonable request to the corresponding authors.

\begin{appendix}
\renewcommand\thefigure{\Alph{section}\arabic{figure}} 

\section{The influence of physics related to magnetic braking on the period gap}\label{Appenx1}
In this section, we propose a possible solution to mitigate the period gap discrepancy in the FW+BDML model.

The condition for magnetic braking to operate varies among different MESA versions. In relatively older MESA versions (e.g. version 11701, as used in this paper), the code checks only the size of the radiative core to determine whether magnetic braking is activated. However, in more recent versions (e.g. version 24.08.1), the code simultaneously check the sizes of both radiative core and convective envelope. Figure\,\ref{Fig-period_gap_diff_MESA} demonstrates that the period gap is influenced by this specific condition used to activate magnetic braking. We see that the edge of the period gap shifts to larger value for MESA version 11701 (blue line) in comparison to version 24.08.1 (orange line). 

However, neither the standard MESA 11701 nor 24.08.1 can perfectly reproduce the period gap with standard magnetic braking alone, unless a more inefficient magnetic braking \citep[e.g. a scaling factor of $f_{\rm mb}$=0.66 for standard magnetic braking, ][]{2011ApJS..194...28K} is applied in version 24.08.1 (green line). In summary, the edge of the period gap is influenced by both the conditions under which magnetic braking operates and its strength. Therefore, in principle, it is possible that the difference for the period gap between the FW+BDML model and observation can be alleviated by adjusting the magnetic braking. 

\setcounter{figure}{0}
\begin{figure*}[ht]
\centering
\includegraphics[width=8cm,height=8cm]{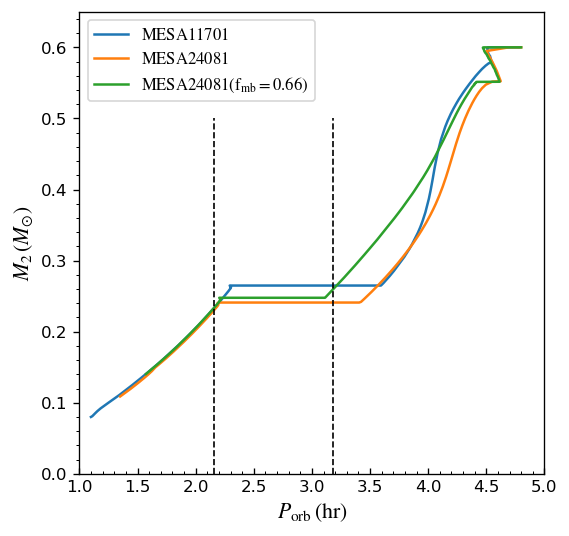}
 \caption{The impact of different conditions under which magnetic braking operates and its strength on the period gap. The blue line and orange line represent the evolution simulated by using standard MESA version 11701 and 24.08.1, respectively. The green line depicts the evolution when a scaling factor ($f_{\rm mb}$) of 0.66 for standard magnetic braking is applied in version 24.08.1. The dotted lines are the edges of period gap (2.15-3.18\,hr; \citealt{2011ApJS..194...28K})}
 \label{Fig-period_gap_diff_MESA}
\end{figure*}


\end{appendix}

\end{document}